\documentclass[10pt,preprint2]{aastex}

\newcommand{\et}{{et~al.\,\,}}

\begin{document}
\title{Giant Pulses from the Millisecond Pulsar B1821$-$24}
\vspace{-0.4cm}

\author{Roger W. Romani\altaffilmark{1,2} and Simon Johnston\altaffilmark{3}}

\altaffiltext{1}{Dept. of Physics, Stanford University, Stanford CA  94305-4060}
\altaffiltext{2}{Visiting Scholar School of Physics, University of Sydney 
\& ATNF, CSIRO, Epping, NSW}
\altaffiltext{3}{School of Physics, University of Sydney, NSW 2006, Australia}

\begin{abstract}

	We have carried out a survey for `giant pulses' in 5 millisecond 
pulsars.  We detect individual pulses from the high ${\dot E}$ 
pulsar PSR B1821$-$24 with energies exceeding 50$\times$ the mean
pulse energy. These giant pulses are concentrated in a narrow phase window
coincident with the power-law non-thermal pulse seen in hard X-rays. This 
is the third example of the giant pulse phenomenon. It supports the 
idea that large $B$ fields in the outer magnetosphere are critical to 
the formation of such pulses and further suggests a direct connection between
giant pulses and high energy emission. 
	
\end{abstract}


\section{Introduction}

	Despite three decades of intensive study, the mechanism producing 
pulsar radio emission is poorly understood. Fluctuations in the intensity
of the radio radiation provide important constraints on plausible mechanisms.
Single-pulse studies of bright pulsars detect a variety of
patterns in the intrinsic intensity fluctuations, including nulling
and drifting phenomena.  The distribution of integrated pulse energies, 
however, has only a modest dispersion, typically forming a Gaussian of width
$\la  \langle E \rangle$, the mean pulse energy (eg. Manchester \& Taylor 1974).
In contrast, the Crab pulsar PSR B0531+24 emits pulses with energies
$> 20 \times \langle E \rangle$, extending up to $> 2 \times 10^3
\langle E \rangle$ (Lundgren \et 1995) which were instrumental in 
the original detection of the Crab (Staelin \& Reifenstein 1968).
These Giant Pulses (GP) are typically broadband (Moffett 1997, Sallmen \et 1999)
and of short duration, with widths of order a few $\mu$s and structure 
down to 10ns (Hankins 2000). They are localized to the main and 
interpulse phase windows and follow an intensity distribution best 
characterized as a power law with index of $\approx 3-3.5$.

	The discovery of similar pulses from the millisecond pulsar
PSR B1937+214 (Sallmen \& Backer 1995, Cognard \et 1996) was surprising.
The pulses are extremely short ($\tau < 0.3 \mu$s) events
confined to small phase windows trailing the main and inter- pulses,
again with an approximately power-law distribution of pulse energies
(Kinkhabwala \& Thorsett 2000). Since PSR B1937+214 is the only known
radio pulsar having an estimated magnetic field at the 
light cylinder larger than that of the Crab, it has been suggested 
that this is a key parameter controlling the giant pulse phenomenon 
(Cognard \et 1996).

	Johnston \et (2001) have recently found that the Vela pulsar
produces pulses with a very wide distribution of peak fluxes. It is not
clear if these events are related to true giant pulses; the largest
such pulses observed to date have $ E < 10\langle E \rangle$, but
these narrow pulses have peak fluxes exceeding $40\times$ the integrated peak
intensity. Relatively few individual pulses have been examined, and
the extended tail in the $E$ distribution of these `giant $\mu$-pulses' 
may continue into the true GP regime.
The pulses are localized to a phase
window off of the integrated pulse emission, are of short duration 
and are highly polarized. 

	To explore the connection between the Crab and PSR B1937+214 GPs
and the large individual pulses of $1-3 \times 10^4$ yr
Vela-like pulsars, we obtained fast time sample data for several
young and millisecond pulsars. We report here evidence for 
giant pulses in the MSP with the next highest known light cylinder field,
PSR B1821$-$24, along with limits on GP emission in other
high-field MSPs. In a companion paper (S. Johnston \& R.W. Romani 2001,
in preparation) we 
discuss large individual pulse energies in young high field pulsars.

\section{Observations}

	Observations were made on 20-22 May 2001, with the Parkes 64-m 
radio telescope. We used the center beam of the 21-cm multi-beam system 
at an observing frequency of 1517.75 MHz. The receiver has a system equivalent 
flux density of 30 Jy on cold sky. The back-end consisted of a filterbank 
containing 512 channels per polarization each of width 0.5 MHz for 
a total bandwidth of 256 MHz.  The polarization pairs are summed, each 
output is then sampled at 80 $\mu$s, one-bit digitized,
and written to DLT for off-line analysis.

	The data were then de-dispersed at the pulsar's nominal dispersion 
measure. The mean and rms of groups of 8192 samples (0.65 s) were examined 
and those which showed obvious signs of interference were discarded.
The data could then be folded synchronously with the pulsar's topocentric 
period to produce a pulse profile.  Our nominal 5-$\sigma$ sensitivity in 
80 $\mu$s is 1.3 Jy. In practice, even after clipping, we experienced 
substantially larger background fluctuations and the rms exceeded our 
expected rms by a factor of 1.5-3. 

	To confirm that we could detect conventional GPs we made short
observations of the Crab pulsar, during which GPs were detected with high
significance. Short integrations with the Vela pulsar off-axis, producing
an effective continuum flux of only 0.6 mJy, confirmed that we could detect 
the largest individual Vela-type giant $\mu$-pulses at these faint 
continuum flux levels. Finally,
and most relevant to the present study we observed PSR B1937+214, obtaining 
1784\,s of integration ($1.15 \times 10^6$ pulses). Our sampling provides
only 19 bins across the pulse profile, but we clearly detect the GP 
distributions in both the main and interpulse components. In both components
the giant pulse peaks occur approximately one bin 
after the corresponding peak of 
the integrated pulse profile. The largest main component GP obtained
had an energy 445 Jy$\mu$s, the third largest had 230 Jy$\mu$s. The phasing
and intensity distribution are well matched to the 1.4 GHz results reported
by Kinkhabwala \& Thorsett (2000).

\begin{figure}[!h]
\includegraphics[scale=0.4]{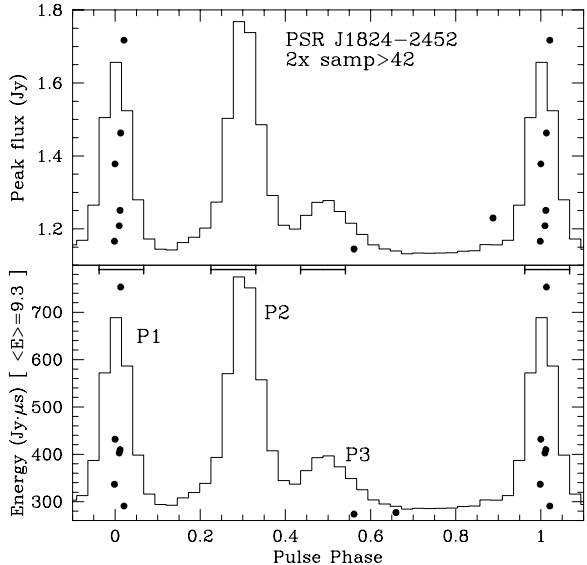}
\caption{Giant pulses from PSR B1821$-$24. Upper panel: the peak flux
of the brightest pulses with adjacent bins $> 4 \sigma$, points at the peak 
phase. Our integrated profile (arbitrary scale) is overplotted for
reference; 1.2 periods are shown.
Lower panel: Integrated energy of the brightest GP.
We label the components on the reference
pulse profile and mark (at top) the 4-bin
windows used to measure component energies.
}
\label{MSP1}
\end{figure}

\section{Results}

	PSR B1821$-$24 has a period of 3.05\,ms and a dispersion measure (DM) 
of $119.8$ cm$^{-3}$pc. This DM leads to a smearing of 142 $\mu$s across our
0.5 MHz filter bands, limiting our resolution and ensuring that any
giant pulse from the source will be smeared across 2-3 time
samples. We obtained 3$\times$3600\,s of data on this pulsar starting
at MJD 52049.7; after excision of 0.65\,s averages showing strong interference,
we retained 9943\,s of data or $\approx$3.3$\times 10^6$ pulses.
Because of the expected DM smearing, we first searched the de-dispersed 
data stream by tagging pulses with 2 or more consecutive samples 
exceeding $4 \sigma$. Our data contained 16 such pulses.
The brightest of these, both in terms of peak sample intensity and
integrated energy were confined to a narrow window in the first component
of the integrated pulse profile (Figure 1, which shows the brightest 8 pulses
by each measure).  During these observations
PSR B1821$-$24 had a continuum flux of $\approx$3 mJy and an average pulse
energy of 9.3\,Jy\,$\mu$s. The largest individual pulse observed had
$E \approx 755 {\rm Jy \mu s}$ which is $81 \langle E \rangle$.

	This pulsar has a complex pulse morphology. We adopt the naming
convention of Backer \& Sallmen (1997), in which P1 has a steep spectrum,
dominating the pulse at low radio frequencies and the second narrow pulse P2
is stronger at short wavelength. The broad trailing component P3 seems to
have a similar spectrum to P1. Backer \& Sallmen report that at 1.4 GHz the 
pulsar occasionally spends many hours in a mode with P2 decreased relative to
the other components by a factor of 4. We do not see such dramatic variations,
but confirm that P2 is variable, decreasing in flux by 25\% relative to 
P1 and P3 over our first hour of integration and stabilizing afterwards.

\begin{figure}[!h]
\includegraphics[scale=0.4]{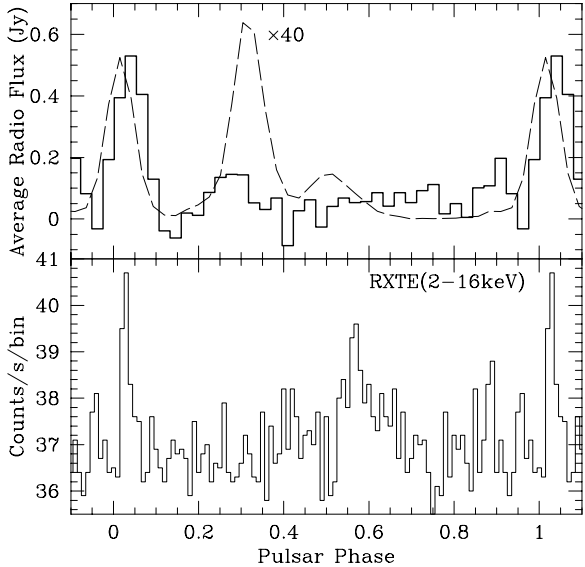}
\caption{Above: the average profile of the 16 GP candidates
(full line histogram) compared with the average profile of all pulses,
magnified by $40\times$ (dashed line curve).
Below: the {\it RXTE} pulse profile with the relative phase determined
by Rots \et (1998).
}
\label{MSP2}
\end{figure}

	The pulsar is also detected in the X-ray, again with a complex
pulse profile. Careful timing (Rots \et 1998) using the
{\it Rossi X-ray Tming Explorer (RXTE)} establishes that 
the narrow, hard spectrum X-ray component is associated with P1, 
but lags significantly by $60\pm 20\mu$s. 
A second broad, and apparently softer, X-ray peak lags P3 by $250\pm 60\mu$s.
Figure 2 compares our integrated pulse profile and the profile obtained
by averaging all 16 pulses with adjacent samples $>4\sigma$. 
These pulses have been re-sampled to common phase bins, with 38 points across 
the light curve; each bin corresponds to about one original sample.
The `GP' lightcurve P1 emission lags that of the integrated pulse profile by
approximately $80\mu$s (one bin). No other significant GP feature is seen. 
For comparison
we plot the {\it RXTE} light curve of Rots \et , with the relative phasing
determined in that study. DM smearing and re-sampling to fixed phase windows
can fully account for our GP pulse width. Thus our GP emission appears to
be unresolved in phase and coincident with the hard power law high energy 
pulse.

We can obtain some further information on the GP distribution
by measuring pulse intensity distributions in windows of pulse phase. For
each pulse we defined a 10 sample off pulse window ($\phi$ = 0.63-0.85
in Figure 1).
The individual components were estimated by the three 4-sample windows 
indicated in Figure 1 (P1=$-$0.03 -- 0.05; P2=0.24 -- 0.32; P3= 0.45 -- 0.53).
For each pulse we determined the background level from the median 
of the samples in off pulse region of the current and preceding pulses.
The individual component energy was measured by an integration of the
energy in the appropriate four sample window. The cumulative histograms 
of the three components are shown in Figure 3. Component P1 shows an
appreciable excess of high energy pulses, extending somewhat below
our selected giant pulse threshold. Note that due to DM smearing and our
limited resolution, some pulse energy will be lost from these 4-sample windows,
so these $E$ values underestimate the full pulse energy.

\begin{figure}[!h]
\includegraphics[scale=0.4]{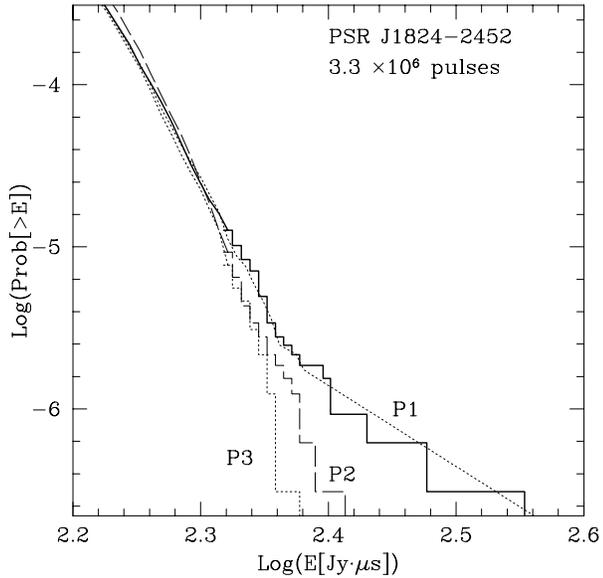}
\caption{Cumulative probability distributions of pulse energy in our 
three selected pulse component windows. The probability histograms are continued
as smooth lines at low pulse energies. There P2 is the highest,
as expected from the average pulse profile. At large energies there is
a clear excess in component P1 (full line histogram). 
A simple power law for the excess GP distribution
is plotted with the dashed line.
}
\label{MSP3}
\end{figure}

	Since only our brightest GPs are well separated from the underlying 
pulse and noise fluctuations, we can put little constraint on their intensity 
distribution.  Power law indices of $\approx 3-5$ adequately describe the
GP excess. Indeed an addition
of an $E^{-5}$ power law to the P3 flux distribution nominally matches
the P1 distribution quite well (dashed smooth line). As for the 1.4 GHz
PSR B1937+214 data of Kinkhabwala \& Thorsett (2000), it is likely
that noise fluctuations and normal pulses steepen the estimated power law
index this close to the noise floor. Longer observations with greater 
sensitivity will be needed to probe the true GP intensity distribution.
We expect about $1/$h ($P \approx 8.5 \times 10^{-7}$) with 
$E > 28 \langle E \rangle$. This 1.4 GHz GP rate is, in fact, 
roughly twice that of PSR B1937+214, which has $P(28\langle E \rangle )
\approx 4 \times 10^{-7}$.  
\begin{tabular}[b!]{lrrrrrr}
\\
\multicolumn{7}{c}{Table 1. MSP Giant Pulse Candidates}\\
\hline\hline\
& P &$B_{\rm LC}$& DM& $\langle E \rangle$& $N_p$ & $E_{\rm lim}$\\
Name & (ms) &($\times 10^5$G)& (${\rm cm^{-3}pc}$)& 
(Jy\,$\mu$s)& $(\times 10^6)$ & ($\langle E \rangle$)\\
\hline
B1821$-$2452 & 3.1 & 7.2 & 120 & 9.3 & 3.3 & $^\dagger$20-80 \\
J0034$-$0534 & 1.9  &  1.4&  14 & 0.1 & 1.9  & 2000 \\ 
J0613$-$0200 & 3.1  & 0.55 & 39 & 9.2 & 4.6  &   27 \\
J2129$-$5721 & 3.7  & 0.50 & 32 & 1.9 & 2.9  &  155 \\
J1911$-$1114 & 3.6  & 0.44 & 31 & 2.6 & 1.5  &  105 \\
\hline
\multicolumn{7}{r}{$^\dagger$ GP detections}\\
\end{tabular}

\noindent
If our GP are
as narrow as those of PSR 1937+214, 
we expect the brightest pulse in an hour
to reach a peak flux of $\ga 1300$ Jy.

\section{Limits for other MSPs}

	We also observed four other millisecond pulsars with high
inferred light cylinder magnetic fields, $B_{\rm LC}$. 
In Table 1, in $B_{\rm LC}$ order, the pulsars that 
we observed are listed along with the mean pulse energy, number of pulses 
observed and the limit to which giant pulses were not seen. These four 
pulsars have $B_{\rm LC}$s substantially lower than that of PSR B1821$-$24.
For PSRs J0034$-$0534, J2129$-$5721 and J1911$-$1114, the low flux
at our observation epoch precluded significant bounds on GP emission. 
The last entry of the table lists the factor by which a pulse must exceed
$\langle E \rangle$ to be detected in our study.  Indeed, if the cumulative 
intensity distribution followed ${\rm Prob}(>E) \propto E^{-3}$ and the GP 
rate were the same as for B1821$-$24, we would have required $10^5$, 58 
and 18 million pulses, respectively, to see a giant pulse in these three 
pulsars at our achieved threshold. Considering further the likely 
decreased GP frequency at lower $B_{\rm LC}$, our limits do not
significantly constrain GP emission in these sources. On the other hand,
for J0613$-$0200 (with a similar period and flux density to
PSR B1821$-$24 but $B_{\rm LC}$ lower by 5), our limits
do begin to probe B1821$-$24 -like GPs. For the GP distribution assumed above
with the same probability as in B1821$-$24, we would have expected about 15
GP events. In fact, the largest few pulses do occur in $0.1$ of phase
ahead of the integrated pulse profile, suggesting that more sensitive 
observations might uncover a GP distribution in this source at $\la 1/10$
the frequency of the events seen in B1821$-$24.
\vfill\eject
	
\section{The Giant Pulse Phenomenon and High-Energy Emission}

	The addition of B1821$-$24 to the GP family supports the
idea that large magnetic fields in the outer magnetosphere are required for
this behavior. However this object points to an additional common feature:
as for the Crab pulsar, the GPs appear isolated to narrow windows of pulse
phase showing high energy pulses, with hard power law spectra. In
the case of the Crab these X- and $\gamma$-ray pulses are believed to
arise in an outer magnetosphere acceleration gap along the 
boundary of
the open zone (Cheng, Ho \& Ruderman 1986; Romani \& Yadigaroglu 1995). 
These Crab-type outer gaps are believed to have $\gamma-\gamma$ pair
production maintained by a dense bath of target soft synchrotron 
photons
from the gap itself. High $B$ fields (especially near the null charge surface)
should enhance this synchrotron emissivity, and hence the rate of
secondary $e^\pm$ production. This dense pair plasma produces the narrow
hard X-ray pulses, and we can speculate that the high densities promote
the instabilities that create enhancements in the particle coherence
and hence the giant radio pulses.

	This picture provides a few predictions. PSR B1937+214 shows
narrow X-ray pulses in ASCA data (Takahashi \et 2001), with a phase close
to the radio interpulse. We expect that a high
resolution X-ray profile with good absolute radio phase comparison will
show these X-ray pulses to align with the GP phase, lagging the integrated
radio profile pulse peak. It is not clear why the interpulse should be 
stronger at high energies than the main pulse.
An additional strong GP candidate not accessible to our survey is the
millisecond pulsar PSR J0218+4232 ($B_{\rm LC} = 3.1 \times 10^5$ G). 
Comparison with the high energy pulse profile (Kuiper \et 2000)
suggests that GP emission may be found
in the second radio pulse, and possibly in the $>100$ MeV pulse phase
preceding the first radio peak. 

	The detailed dependence of the GP frequency
on $B_{\rm LC}$ is not clear. If the critical quantity is the field
near the outer gap base at the null charge surface, where pair
production is expected to be most intense, then magnetic inclination $\alpha$ 
can have a strong effect with $B \approx B_{\rm LC} [1.5\, {\rm tan}(\alpha ) ]^6$
for a static dipole field and $\alpha \ga 30^\circ$. Consider the case of the
Crab pulsar, which has a frequency of $E > 20 \langle E \rangle$ pulses
roughly  $10^4 \times $ larger than that of
PSR B1937+214 or PSR B1821$-$24 at 1.4 GHz. Polarization modeling
(Romani and Yadigaroglu 1995) and the recent CXO wind torus images 
suggest that the Crab has $\alpha \approx 80^\circ$. Baker and Sallmen (1997)
found a plausible polarization sweep fit of $\alpha \approx 50^\circ$ 
for PSR B1821$-$24. For this angle the nominal field at the base of the gap
is suggestively $10^4 \times$ lower than that of the Crab. Of
course, the magnetosphere of a MSP is likely far from vacuum dipole,
but this does underline that $\alpha$ can have a strong effect on the 
ordering of GP candidates by $B$.

	A better criterion for the presence of strong GP activity may be 
the presence of a narrow hard spectrum X-ray pulse, indicating a dense pair 
plasma in an active outer gap. Most known, young $\gamma$-ray pulsars
are Vela-like with pair production controlled by thermal surface fluxes
(e.g. Romani 1996).  These do have non-thermal X-ray pulse components, but
these are much broader and the emission represents a much smaller
fraction of the spindown luminosity. GP distributions might therefore be
much weaker in these pulsars.  The occurrence of GP in other young Crab-like 
pulsars and the connection with the large pulses from millisecond
pulsars described here can only be addressed by further sensitive single 
pulse studies.

\begin{acknowledgements}
The Parkes telescope is funded by the Commonwealth of Australia
for operation as a National Facility managed by CSIRO. 
The 512 channel filterbank system used in the observations
was designed and built at the Jodrell Bank Observatory, University 
of Manchester.  Observing software was provided by the Pulsar 
Multibeam Survey group.

\end{acknowledgements}


\end{document}